\documentclass[preprint,3p,11pt]{elsarticle}

\usepackage{amssymb}
\usepackage{color}
\usepackage{bm}
\usepackage{url}

\usepackage{natbib}

\journal{COCIS}

\begin{document}

\begin{frontmatter}

\title{Leveraging Collective Effects in Externally Driven Colloidal Suspensions: Experiments and Simulations}

\author[label1]{Michelle Driscoll\corref{cor1}}
\address[label1]{Department of Physics and Astronomy, Northwestern University, 2145 Sheridan Rd, Evanston, IL 60208-3112}
\cortext[cor1]{Corresponding author.}
\ead{michelle.driscoll@northwestern.edu}

\author[label2]{Blaise Delmotte}
\address[label2]{LadHyX, UMR CNRS 7646, \'Ecole Polytechnique, 91128 Palaiseau CEDEX, France}
\ead{blaise.delmotte@gmail.com}

\begin{abstract}
In this review article, we focus on collective motion in externally driven colloidal suspensions, as well as how these collective effects can be harnessed for use in microfluidic applications. We  highlight the leading role of hydrodynamic interactions in the self-assembly, emergent behavior, transport, and mixing properties of colloidal suspensions.  A special emphasis is given to recent numerical methods to simulate driven colloidal suspensions at large scales. In combination with experiments, they help us to understand emergent dynamics and to identify control parameters for both individual and collective motion in colloidal suspensions. 
\end{abstract}


\end{frontmatter}

\section{Introduction}

At equilibrium, colloidal suspensions can organize into familiar thermodynamic phases (gas, liquid, crystal, glass,\ldots), and our statistical mechanics toolkit has been quite helpful in understanding their physics \cite{Anderson2002}.  However, many of the suspensions we interact with in our everyday life are not at equilibrium, but flowing.  The out-of-equilibrium nature of a flowing suspension leads to new and rich behavior such as  complex rheology, and the emergence of new kinds of ordering and structure.  Here we will discuss suspensions which are \emph{externally driven} by an applied field.  In these systems, an  external field applies  force and/or torque to individual particles.  Seemingly, this would lead to quite limited possibilities for self-organization --- but due to strong collective interactions between particles, complex structures and non-trivial behavior can emerge.

This review focuses on the  behavior of driven colloidal suspensions, a subject that has received less attention than understanding the dynamics of  individual driven/active particles.  There has been an explosion of experimental techniques to manufacture new kinds of colloids, and several recent reviews focus on understanding their propulsion mechanisms from a single particle level \cite{Aubret2017, Xu2017b, Martinez2018, Zottl2016, Zhang2015}. 
Here, we will discuss systems in which particles are driven with an external field, in contrast to those driven by local input of chemical energy. For a thorough discussion of chemically-driven systems, we guide to reader to the following reviews \cite{Aubret2017, Liu2017,  Ebbens2010}.

Our aim is to highlight recent work in the field;  more complete reviews exist which discuss each of the driving strategies we mention  in greater technical detail.  We have organized this review by the type of external drive: magnetic, electric, and acoustic. For the sake of conciseness we do not cover collective motion in suspensions of optically driven colloids. We refer the reader to pioneering and more recent studies on hydrodynamic synchronization and clustering in optically driven systems \cite{Lutz2004,Roichman2007,Kotar2010,Sokolov2011,Curran2012,Nagar2014,Okubo2015} and to recent reviews on the subject  \cite{Bowman2013,Sukhov2017,Polimeno2018,Martinez2018,Saito2018}.  Here, we have focused on assembly and structures with an eye towards transport and microfluidic applications; we do not cover the large body of experiments focused on self-assembled static structures, and recommend the following reviews for more information \cite{Yi2013,Yang2016,Manoharan2015, Vogel2015, Frenkel2014}.

For each of the driving strategies, we emphasize self-assembly or spatio-temporal organization which occurs due to collective effects. We underline the crucial role of long-ranged hydrodynamic interactions, which, together with externally induced interactions, lead to these strong collective effects. In such situations,  theoretical calculations can only handle one or two particles at most. Numerical simulations thus represent the most appropriate tool to study many-body systems such as colloidal suspensions. We have devoted attention to highlighting those simulation techniques and their recent advances.  This is a rapidly emerging field; it is now possible to simulate not just a handful of single particles, but high-density suspensions with $O(10^6)$ particles with a  reasonable amount of computational power.  Coupled with experiment, detailed simulations can help us to understand the complexity of structures which emerge from collective dynamics in driven suspensions.   We conclude by highlighting promising future directions and open questions.

\section{Magnetically powered colloidal suspensions}
External magnetic fields offer a promising avenue for controlling dense colloidal suspensions.  Low powered magnetic fields are relatively easy to generate in the lab, and offer the possibility of three dimensional control of particle orientation.  Unlike chemical or electrical driving mechanisms, magnetic driving has the advantage of being inherently bio-compatible, opening up exciting new avenues for application possibilities.  We briefly outline magnetic driving  below; for a more detailed description, we refer the reader to recent reviews focused on magnetic propulsion \cite{Han2017, Martinez2018, Snezhko2016, Rikken2014}. 

\subsection{Particle-based driving}

The most straightforward way to use an external magnetic field to drive a colloidal suspension is to use the field to directly manipulate individual colloidal particles.  The field is used to apply a torque to individual particles, which can then be converted into translational motion using a variety of strategies.

The appropriate orientation of the magnetic field depends on the relative balance of viscous and inertial forces.  In the overdamped limit, where inertia can be neglected (very small Reynolds number), the applied torque must result in a particle motion that breaks the time-reversal symmetry inherent in Stokes flow.  One strategy for generating this motion is to mimic bacteria by creating an artificial flagella or cilia, which is then actuated in either a  travelling-wave or helical motion. This non-reciprocal actuation breaks the time-reversal symmetry and generates propulsion \cite{Martinez2018, Peyer2013}.  

Alternatively, one can use the magnetic field to rotate  particles near or on a solid surface, so that either the surface forces or the particle-surface hydrodynamic coupling generate motion.   If the axis of rotation is oriented parallel to a surface, rotation leads to linear translation of the particle  in both cases \cite{Dean1963,Goldman1967}.  We note that both permanently magnetized as well as ``super-paramagnetic'' colloids can be manipulated in this way \cite{Martinez2018}.

At small but finite Reynolds number, a \emph{vertically oscillating} magnetic field can be used to generate particle rotation.  Ferromagnetic colloids will rotate either clockwise or counterclockwise, following the applied field.  If the particles are large enough, the lag generated by inertia can lead to spontaneous symmetry breaking, so that individual particles begin to continuously roll \cite{Kokot2015}.  While this symmetry breaking mechanism is reminiscent of Quincke rotation (see discussion in section \ref{electric}), we note two important differences: rotation speed is independent of field strength, and the particle dynamics are intrinsically chaotic  \cite{Kaiser2017}.  

\subsection{Substrate based-driving}

An alternative to particle-level driving is to instead use a patterned magnetic substrate.  By applying a time-varying magnetic field to the substrate, one can alter its magnetic properties in a such a way as to drive adjacent particles.  One strategy, employed by Yellen et al., is to immerse non-magnetic particles in a ferrofluid above the patterned substrate \cite{Yellen2005, Zhu2011}.   The nonmagnetic particles then experience a force governed by the amount of magnetic fluid they displace, and by using a time-or-spatially varying magnetic field to adjust the magnetization of the patterned surface particles can be transported at rather high velocities ($\sim 70 \mu$m/s).  While there are some limitations due to immersion in a ferrofluid, this system is promising as it allows the transport of a wide variety of particles.

An alternative strategy employs uniform ferromagnetic films with patterned magnetic domains.  These  magnetic domains can be organized into a variety of structures such as stripes or bubbles.  These structures (Bloch walls) can then be manipulated with an external magnetic field; adjacent magnetic particles  couple to the  domain motion and translate \cite{Tierno2009, Ehresmann2011, Gunnarsson2005}.  An underlying striped lattice confines transport along a single dimension, while using a geometry with more symmetry such as a bubble lattice allows for guided transport along an arbitrary 2D path, see Figure \ref{fig:Magnetic}a.  Patterned magnetic substrates, while limited in their reconfigurability, are a promising system for microfludic transport, as they allow for particle transport at quite high velocities \cite{Tierno2009}.  
Recent work with these substrates has demonstrated that, due underlying to lattice symmetries, certain transport paths are topologically protected \cite{Loehr2017, Loehr2016}.  This system has also been used to create a landscape with quenched disorder by permanently placing large obstacles at random positions on the substrate, and then studying their interaction with translating magnetic particles  \cite{Stoop2017}.  We believe that interaction with non-trivial boundaries is an exciting direction of study, and necessary to understand suspension transport in complex environments.

\subsection{Collective behavior: self assembly via rotating fields}

Magnetically actuated suspensions can self-assemble into a variety of  transient and permanent structures.  One of the most common means of magnetic driving is to apply a rotating field, with an additional degree of control offered by the orientation of the field relative to a nearby surface which influences the particles due to hydrodynamic coupling.  As first shown by experiments with mm-scale spinning disks, the balance between magnetic  and hydrodynamic interactions can lead to the self-assembly of organized lattices \cite{Grzybowski2002}.  By altering the relative strength of magnetic vs.\ hydrodynamic interactions, the collective behavior of the suspension can be dramatically changed.  We note that here we only consider bulk fluid suspensions.  When a fluid-fluid interface is present, quite different dynamics control the collective behavior, which are discussed in depth in the recent review by Snezhko \cite{Snezhko2016}.

\begin{figure}
    \centering
    \includegraphics[width=0.95\columnwidth]{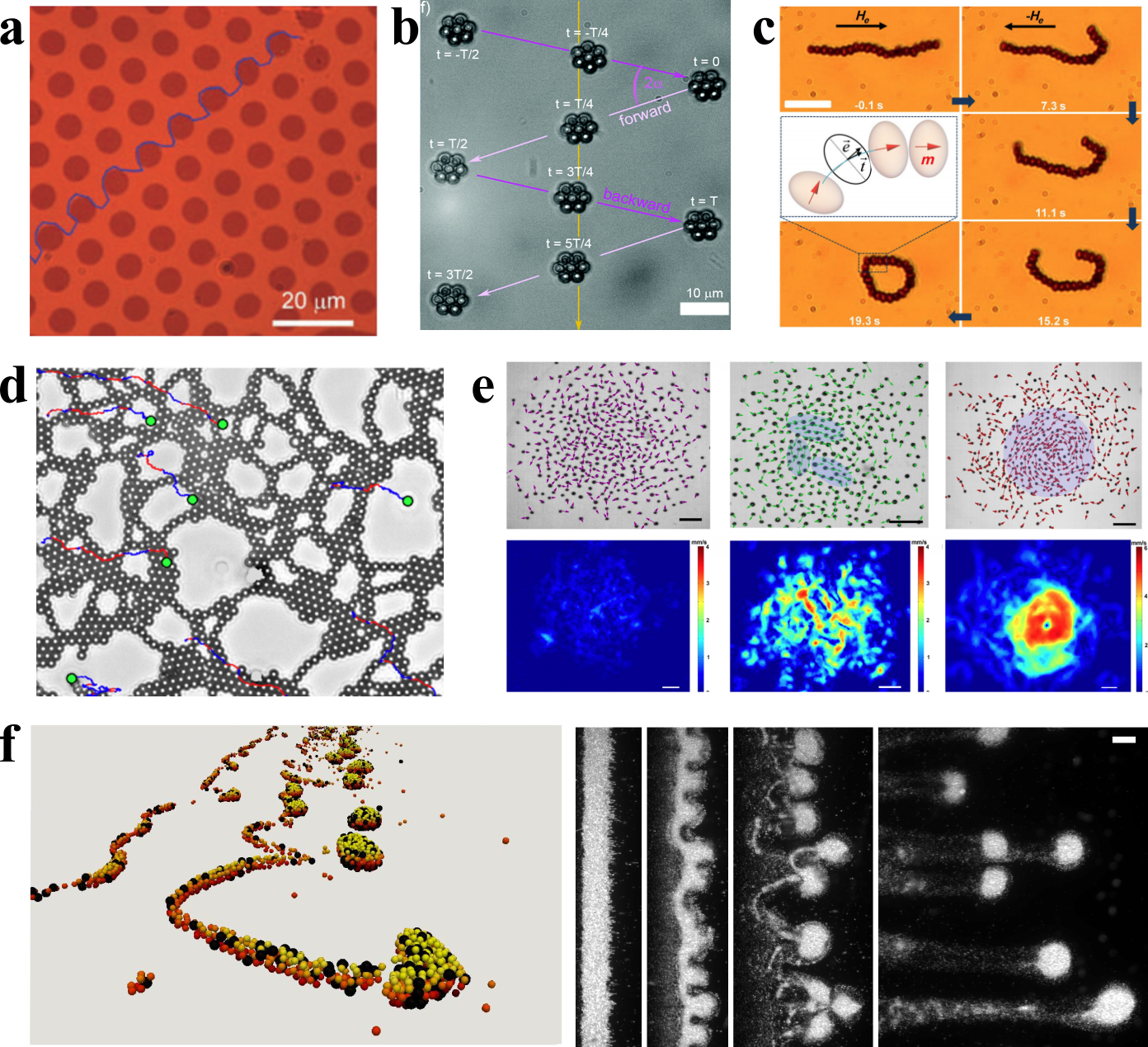}    \caption{Examples of individual and collective motion in magnetically driven colloidal suspensions. \textbf{a}, Transport of a particle using a patterned magnetic substrate, adapted from \cite{Tierno2014} \textbf{b}, Guided transport of a self-assembled colloidal wheel, adpated from  \cite{Maier2016b}. \textbf{c}, A flexible magnetic chain that is externally manipulated to form a ring, adpated from \cite{Martinez2016a}. \textbf{d} Transport of passive (non-magnetic) particles on a network of actively rotating (but not translating) 1.4 micron magnetic colloids, adpated from \cite{Maier2016a}.  \textbf{e}, Asychronously rotating colloids dynamically assemble into a rich variety of phases, shown from left to right are experimental images of the gas, flocking, and vortex phase.  The lower images represent course-grained velocity measurements (scale bars are 1 mm), adapted from \cite{Kaiser2017}.  \textbf{f}, Creation of hydrodynamically stabilized clusters (critters) from a hydrodynamic fingering instability in both simulation (left) and experiment (right).   The particles are rotated near (but not at) a wall; the particle-wall distance sets the scale of the instability as well as the size of the critters (scale bar is 200 micron). } 
    \label{fig:Magnetic}
\end{figure}

In the limit of strong magnetic interactions, the inter-particle attraction due to the spinning magnetic dipoles can be used to self-assemble a variety of dynamic structures such as rotating crystals, chains, tubes, and ``colloidal wheels'' \cite{Tasci2016, Maier2016b, Yan2015a, Yan2015b}.  Colloidal wheels represent an exciting new development; they translate at quite high velocities, up to 50 $\mu$m/s, and have unique dynamics due to the fact they assemble and roll with a slight tilt, see Figure \ref{fig:Magnetic}b. Recent work has demonstrated that, unlike a bicycle wheel, this dynamic structure is chiral; its motion does not reverse upon reversal of the driving field.  Thus, this system offers a minimal experimental model for studying the necessary ingredients for non-reciprocal motion \cite{Maier2016b}.

 A particularly rich set of dynamics emerges in a system of asynchronously rotating ferromagnetic colloids. Here, large (50 micron) colloidal particles are driven by a uniform vertically oscillating field, but the rotation direction of individual particles is defined by spontaneous symmetry breaking, so that it can be altered due to collisions \cite{Kokot2015}.  Hydrodynamic interactions are very weak, and particles assemble into structures dictated by magnetic interactions and interparticle collisions.  By sweeping the frequency of the applied field, the particles in this system assemble into several correlated phases, exhibiting both vortical motion and flocking, see Figure \ref{fig:Magnetic}e \cite{Kaiser2017}. This system holds great potential for studying flocking transitions in more complex systems, including living matter. Recent work by Kokot et al.\ has demonstrated the existence of an active vortex phase \emph{without} the need for geometric confinement, and explored its interaction with passive particles \cite{Kokot2018}.  This system shows promise for providing insight into how obstacles can be used to manipulate and guide self-assembled structures, and offers a new route to active particle transport.

In a system of synchronous rotating magnetic colloids, the structures which self assemble are again governed by the balance between magnetic and hydrodynamic interactions.  When magnetic interactions between the particles are strong (colloids with large magnetic moment), particles assemble into chains, crystalline structures, and networks \cite{Martinez2015, Martinez2015b, Maier2016a}.  These dense structures can be used to transport cargoes; although the particles are closely packed, they can still follow the applied rotating field, creating a pumping flow that can be used for advective transport, see Figure \ref{fig:Magnetic}d. In the intermediate range, when magnetic interactions are comparable to hydrodynamic interactions, chain-like structures assemble, but are much more flexible.  Due to the finite relaxation time of particle magnetization, these chains orient perpendicular rather than parallel to the applied field \cite{Martinez2016a, Martinez2016b, Massana2017a, Massana2017b}. The hydrodynamic flow adjacent to these assembled chains can be used to shuttle cargo \cite{Massana2017b}. Additionally, these chains can easily be made into rings, which can be assembled around a cargo particle, and then used to transport it, see Figure \ref{fig:Magnetic}c \cite{Martinez2016a, Yang2017b}. 

When magnetic interactions are very weak, so that hydrodynamic coupling is the dominant particle-particle interaction, very different structures emerge.  In this case, although the external magnetic field provides a convenient way to rotate particles, interparticle magnetic interactions can be completely neglected as compared to viscous forces/torques acting on the particle.  In this weakly magnetic case, rotation adjacent to, but a finite distance above, a no-slip surface leads to very strong collective effects.  (This is quite different rotating a particle \emph{at} a surface, which lead to translation akin to a wheel). In this case, a uniform suspension of particles shows density waves which propagate freely throughout the system \cite{Delmotte2017}.  If a large enough density perturbation is introduced into the system, a whole cascade of instabilities is observed: first a shock is formed, and then this shock quickly destabilizes into fingers with dense tips \cite{Delmotte2017b, Driscoll2017}.  The particle-wall distance becomes the dominant lengthscale in the system, controlling both the width of the shock front as well as the wavelength of the instability.  As the instability grows, the dense tips of these fingers can then break off, and continue to travel as persistent, dense clusters termed `critters', see Figure \ref{fig:Magnetic}f.  Critters are bound together only through hydrodynamic interactions; if they are indeed a stable state of the system, this demonstrates that stable clustering does not require attractive interactions, but can be created by hydrodynamic coupling alone.  These kinds of hydrodynamic bound states have also recently  been demonstrated with pairs and chains of elliptical particles \cite{Martinez2018b}.  Hydrodynamic bound states are an important area to explore further, and there are still fundamental questions about the limits of their stability.  From an applications standpoint, critters offer exciting possibilities for encapsulation and transport of cargoes at the microscale.

\subsection{Collective behavior: bio-mimetic systems}

The collective behavior of most bio-mimetic systems has been studied considerably less than that of simpler colloidal particles, likely due to the difficulty of manufacturing magnetically controlled artificial flagella at large scales.  A few studies have been done, which demonstrate that these more complex particle shapes display clustering instabilities reminiscent of those seen in magnetic sphere systems \cite{Vach2017}. This is an exciting area for future work: as larger-scale fabrication of these artificial flagellated systems becomes possible, we expect they will help enlighten open questions on bacterial swarming and flocking.  Particle shape has a strong effect on hydrodynamic interactions, so studying these questions with spherical colloids alone is insufficient.  For a more thorough overview of recent advances in creating anisotropic magnetic colloids, see the recent review by Tierno \cite{Tierno2014}.

Recent advancements in microfabication have made fabrication of cilia-like microstructures possible at a relatively large scale, allowing for the study of large-scale collective behavior.  Large collections of cilia are know to beat synchronously; studying simplified models of cilia both allows one to understand the biological system, as well as to explore coupled-oscillator dynamics \cite{Bruot2016, DiLeonardo2012, Damet2012}.  Self-assembled artificial cilia were first studied experimentally by Vilfan et al., who demonstrated the feasibility of artificial cilia as a platform for creating pumping flow \cite{Vilfan2009}.  Carpets of artificial cilia can also be assembled from individual magnetic rods  \cite{Coq2011}.  The rods in this system are created using a soft-lithography  templating technique to align paramagentic beads.  As the beads are fixed in place, the hydrodynamic interactions are that of many coupled flexible rods.  Through modeling in addition to carefully controlled experiments, this work showed that that the collective beating of cilia is controlled by large-scale hydrodynamic coupling of the entire carpet of artificial cilia. Additional studies have recently succeeded in creating magnetic cilia by robust and simple methods \cite{Zhang2018, Hanasoge2018, Hhanasoge2018b}.  We hope to see more activity in this area in the future, it is highly relevant from both a biological point of view (many tissues contain cilia) and an applications point of view (it is an efficient system for microscale pumping).  

\subsection{Outlook}
Magnetically-driven colloidal suspensions are a powerful tool from an applications point of view: they are inherently bio-compatible and often easily re-configurable.  As techniques to fabricate magnetic particles open up possibilities for particles with more complex shapes, we expect that this will continue to be a very active area of study.  Additional possibilities are opened up by combining magnetic driving with other methods, such as electrical or acoustic driving.

\section{Electrically powered colloidal suspensions}
\label{electric}

Electric driving in colloidal suspensions is more complex than magnetic driving: except at very high frequencies, the suspending fluid as well as the particles can react to the applied field.  Electro-hydrodynamics is a rich subject, and it is not our intent to discuss it in detail.  Rather, we will highlight recent work in this field that illustrates the variety of driving strategies, and discuss emergent behavior due to collective effects. 

\subsection{Electric driving}

Many colloidal suspensions are charge-stabilized, i.e.\ the individual particles have a net charge.  If a homogeneous, DC electric field is applied to a suspension, the charged particles will migrate; this is termed electrophoresis.  Even uncharged dielectric particles can be polarized in an applied field; if a field gradient is present this can also lead to particle motion, termed dielectrophoresis.  Colloids can also be manipulated by using an electric field to act on the (charged) suspending fluid.  Even initially uncharged colloids placed in a uniform field will polarize, and this will result in the particle being surrounded by a cloud of screening ions, termed the electric double layer.  The external electric field  exerts a force on this double layer, resulting in fluid flow.  This effect is termed `induced-charge electrophoresis' or ICEP, as the applied field is exerting a force on the charge which is induced by the field itself.  ICEP generates a quadrapolar flow around a symmetric particle, which will not lead to any translational motion.  However, if a particle has some shape or charge asymmetry these flows become asymmetric and can lead to translation, see Figure \ref{fig:Electric}a. A simple example of this is the charge asymmetry present in a bi-material colloid, such as a janus colloid.  We have only briefly mentioned possible driving strategies,  but we emphasize that understanding electrokinetic effects is a rich and active area of study.  For a more in-depth discussion of electric driving of dielectric/conducting particles, we direct the reader to the following reviews \cite{Anderson1989, Squires2006, Bazant2010, Van2013}.

A very different approach to electric driving is to take advantage of a rotational instability first observed by Quincke, who found that a dielectric colloidal particle in a strong electric field can begin to  spontaneously rotate \cite{Quincke1896}.  This rotation occurs when there is a  difference in the charge relaxation time between the colloid and the fluid.  In a strong electric field, a dielectric particle will have an induced dipole moment.  If the charge relaxation time of the colloid is larger than that of the fluid, this induced dipole will be aligned anti-parallel to the applied field.  This configuration is unstable to infinitesimal perturbations; any small rotational displacement will produce a torque which amplifies the perturbation.  This  results in spontaneous rotation at a constant rate in a direction transverse to the applied field \cite{Melcher1969}.    When these rotating particles are adjacent to, or in contact with, a solid surface, they will translate, see Figure \ref{fig:Electric}b.  We note that the particles in a suspension of `Quincke rollers' may translate in different directions as the rotation axis is only confined to be parallel to the surface. However, collective effects  can synchronize particle motion.

\subsection{Collective behavior}

\begin{figure}
    \centering
    \includegraphics[width=0.95\columnwidth]{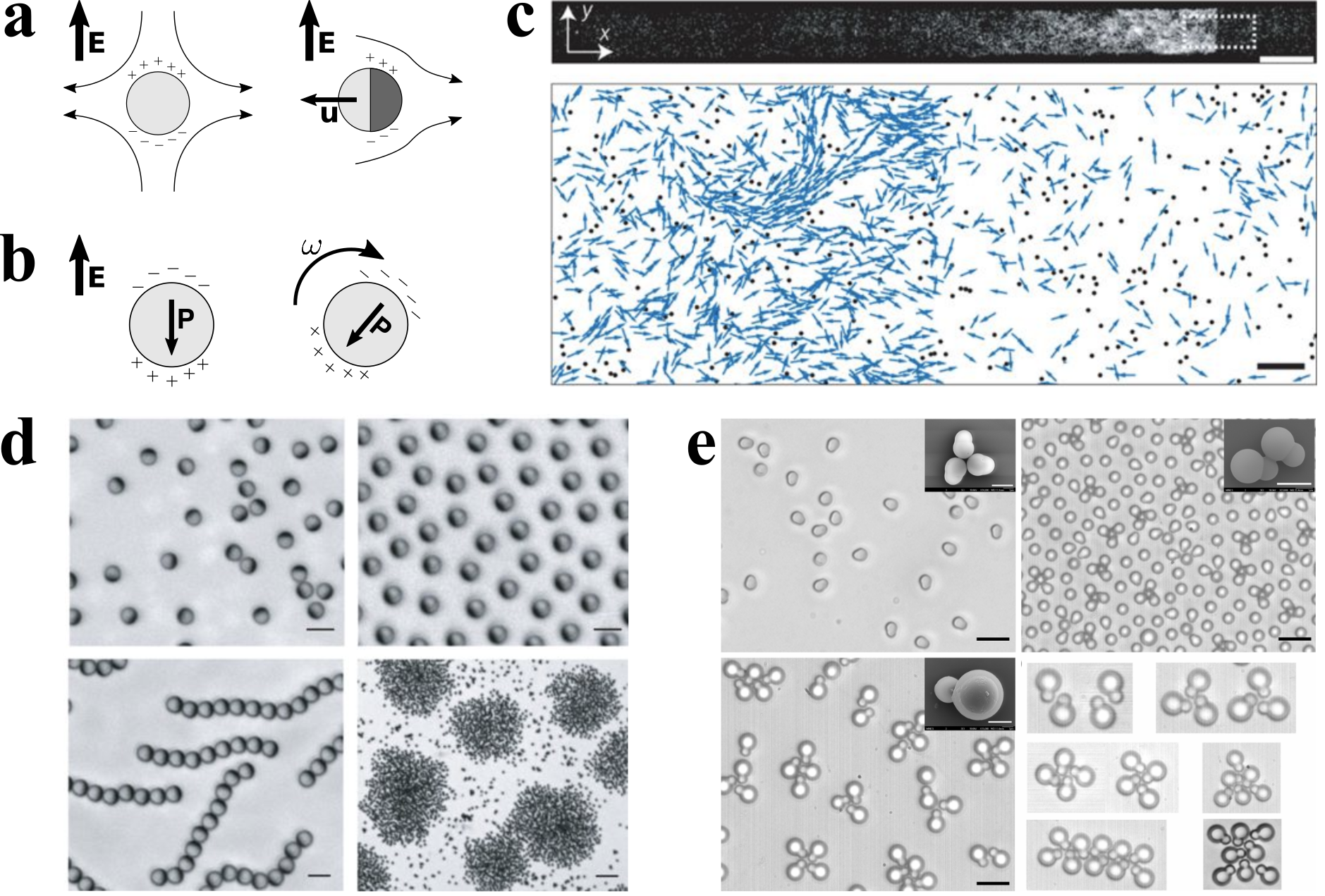}
    \caption{Examples of individual and collective motion in electrically driven colloidal suspensions. \textbf{a}, Sketch of ICEP flow due an applied field, $\bf\vec{E}$.  The flow is quadrapolar around a symmetric particle (left).  However, if charge or shape asymmetry is present, ICEP flows are no longer symmetric and lead to translation at a velocity $\bf{\vec{u}}$ (right). \textbf{b}, Illustration of Quincke rotation.  When the charge relation time is larger in the colloid than in the fluid, the colloid's polarization, $\bf\vec{P}$, will be antiparallel to the applied field, $\bf\vec{E}$ (left).  This unstable configuration results in steady rotation at a frequency $\omega$ (right).  \textbf{c} A flock of Quincke rollers encountering a random lattice of obstacles, adapted from \cite{Morin2017}.  The upper panel shows a zoomed-out view of the colloidal flock (scale bar is 5 mm), and the lower panel shows a close-up of the flock moving past the obstacles, arrows indicate particles and dots indicate obstacles (scale bar is 100 $\mu$m).  \textbf{d} Illustration of different phases achieved by tuning the driving frequency in a system of janus colloids (scale bar is 5 $\mu$m in all bottom left, where it is 30 $\mu$m), adapted from \cite{Yan2016}.  \textbf{e}, Collective behavior of asymmetric particles (colloidal dimers), adapted from \cite{Yang2017}.  By modifying the shape asymmetry as well as the surface charge, a variety of behaviors is observed: gas, 3D clusters, and planar clusters; bottom left panel shows a variety of cluster sizes.  Large images are optical images (scale bars 5 $\mu$m) and small insets are SEM images (scale bars 2$\mu$m).   } 
    \label{fig:Electric}
\end{figure}

As discussed above, there are a large variety of strategies for electrically driving suspensions.  For example, Vissers et al.\ demonstrated that even a simple system of oppositely charged colloids in DC field  can organize into parallel lanes due to collective effects \cite{Vissers2011}.  Though asymmetry is required for  transport via ICEP flows, symmetric particles have been shown to organize in clusters, chains, and organized bands and vortices due to hydrodynamic interactions competing with other effects \cite{Zhang2006, Hu1994, Yeh1997, Fraden1989, Perez2010}.  However, fully leveraging ICEP effects for driving suspensions requires some particle asymmetry.  While new techniques are expanding our ability to create controllable shape asymmetry, charge asymmetry is more commonly implemented.  This is easily achieved by using Janus colloids, and  they remain a popular choice due to their ease of fabrication \cite{Zhang2017}.  Janus colloids driven via ICEP translate at quite high velocities (tens of $\mu$m/s) \cite{Gangwal2008}.  Suspensions of Janus particles have been shown to assemble into a variety of structures, such as chiral clusters, flocks, and chains \cite{Ma2015, Nishiguchi2018, Yan2016}.   An exciting recent development in this field is the work of Yan et al., which demonstrates that by sweeping the field frequency, they could tune the dipolar interactions between individual particles.   By adjusting these electrostatic interactions, they achieved several varieties of ordered phases from one species of particle: chains, swarms, and clusters, see Figure \ref{fig:Electric}d \cite{Yan2016}.   This is an interesting system, as in contrast with assembly in many other suspensions,  the observed phases result from simple pairwise interactions, and not hydrodynamic flows.  As techniques for fabricating Janus colloids become more sophisticated, new possibilities for driven assembly and transport are being realized.  For example, by making a Janus particle that could be translated via ICEP and steered via a magnetic field, Demirrors et al.\ recently demonstrated the potential for guided cargo transport \cite{Demirors2018}.  

Although chemical heterogeneity has been more well-explored, shape asymmetry can also be used to leverage ICEP for particle translation.  Ma et al.\ demonstrated that shape asymmetry could lead to propulsion in experiments using asymmetric dimers.  Additionally, they showed that these flows can be altered by changing the chemical proprieties of the dimers, i.e.\ identically shaped dimers made from different materials can move in different directions \cite{Ma2015b}.  In a suspension of these dimers, the assembly behavior is controlled by the asymmetric ICEP flows.  These flows can be finely tuned by altering particle charge or the conductive properties of the fluid, leading to  the assembly of a variety of chiral and achiral clusters \cite{Yang2017}, see Figure \ref{fig:Electric}e.  As fine control of shape and material properties of colloidal particles is becoming more and more technologically feasible, we hope to see more efforts in this area.  Recent theoretical work by Brooks et al.\ highlights the potential of using shape to create directed motion with ICEP flows.  They present a systematic investigation of dynamics as a function of particle symmetry, demonstrating that essentially any dynamic function (transitional/rotational motion) can be achieved by using an appropriately shaped colloid \cite{Brooks2018}.  We hope this work inspires follow up experimental work, creating driven suspensions with new properties.  It will be especially exciting to see explorations of the collective behavior of these ``programmed'' colloids.

The collective dynamics of Quincke rollers has a rich phase behavior.  Using lateral confinement, experiments have demonstrated that large-scale collective motion can emerge from a uniform population of Quincke rollers .  Polar bands, flocking, and vortex states were all observed.  As both the hydrodynamic and electrostatic particle-particle interactions  are reasonably well-understood in this system, an analytic model of this flocking  transition was constructed in addition to the experimental observations  \cite{Bricard2013, Bricard2015}. Like the asynchronous magnetic system discussed in the previous section \cite{Kokot2018}, confined Quincke rollers are a promising model system for understanding flocking transitions in more complex active matter systems, such as biological ones (bacteria, birds, fish).  Recently, artificial flocks were used to explore the interaction of these dynamic clusters with a disordered landscape, see Figure \ref{fig:Electric}c.  Morin et al.\ found that flocking was suppressed above a critical amount of disorder, suggesting a first order phase transition they argue should occur for all kinds of flocking \cite{Morin2017}.  Additionally, this group explored the response of colloidal flocks to an external field.  They find that although individual particles behave analogously to spins in a magnetic field,  collectively these flocks show a quite different and nonlinear response, with the ability to align and travel opposite the direction of applied flow \cite{Morin2018}.  Mapping out the response to an applied field is important to understanding the stability and dynamics of the structures and condensed phases which arise in driven suspensions, and we look forward to seeing more work along these lines in other systems.  

\subsection{Outlook}

It is only within the last ten years or so that fabrication techniques have become advanced enough to design particles with chemical/shape asymmetry so that ICEP flows can be used for propulsion.  This is a very active area of development, and we expect to see new studies examining particles with complex asymmetry.  Other means of electrical driving, such as Quincke rotation, can be used to create well-understood model systems for exploring flocking.  Very different behaviors  are observed when Quincke rollers are suspended in liquid crystals \cite{Jakli2008} or near fluid surfaces \cite{Ouriemi2014}.  This highlights that in both this and other systems, more work is needed exploring the behavior of driven suspensions near fluid and elastic interfaces.

\section{Acoustically powered colloidal suspensions}

Another way to drive colloidal particles is to use acoustic fields \cite{Lenshof2012,Rao2015,Connacher2018}\footnote{We highly recommend the Acoustofluidics tutorial series of 23 papers in \textit{Lab on a Chip}.}.
Historically, acoustic waves have been used in microfluidics for particle sorting via acoustic levitation. Less than a decade ago, it has been shown that acoustic waves could also be used to propel asymmetric particles with an hydrodynamic effect called acoustic streaming \cite{Wang2012,Nadal2014}, thus opening a new route to study active matter and collective motion.
 In this section we will briefly explain the levitation and propulsion mechanisms and then focus on the collective behaviour induced by acoustic forces and hydrodynamic interactions.

\subsection{Levitation mechanism}

Particles are manipulated by acoustic fields as follows: a vertical standing wave in a reflective channel  generates a force, called the primary radiative force, that drives particles to the pressure nodes or antinodes.  
 To calculate the magnitude of the radiative force,  we consider a one dimensional standing wave in the $z-$direction (see Figure  \ref{fig:Acoustic}a), with  wavenumber $k=2\pi/\lambda$ and acoustic energy $E_{ac}\sim p_0^2$, where $p_0$ is the wave amplitude.  Then, the primary acoustic radiative force in the $z-$direction acting on a spherical particle with radius $a$ is given by \cite{Lenshof2012,Rao2015}
\begin{equation}
    F^{\mbox{rad}} = 4\pi a^3 E_{ac} k \sin(2kd)\Phi.
\end{equation}
Here $\Phi$ is the acoustic contrast factor, and $d$ is the particle distance to the pressure nodes or antinodes. $\Phi$ compares the material properties (compressibility and density) relative to the suspending medium. When $\Phi>0$, which is the case of most colloids, the particle is attracted to the pressure nodes in the acoustic standing wave field. When $\Phi<0$ the particle is attracted to the antinodes. 

\subsection{Acoustic streaming and individual behaviour in the levitation plane}

When a suspension of spherical colloids is placed in a 1D acoustic standing wave along $z$, it will first move towards the pressure nodes, where the primary radiative force is zero, and then follow the acoustic energy distribution in the plane.
When the particle density is different than that of the fluid, the particles oscillates along $z$ with the frequency of the acoustic wave, and the relative motion of the particle with respect to the fluid generates non-linear terms whose period-averaged value is non-zero and results in a steady flow (e.g.\ the nonlinear term $\cos(\omega t)\cos(\omega t) = \frac{1}{2}(1+\cos(\omega t))$ has a non-zero period-averaged value of $\frac{1}{2}$) \cite{Sadhal2012,Sadhal2012b}.  This phenomenon is known as acoustic streaming; Figure  \ref{fig:Acoustic}b illustrates the streaming flow around an oscillating sphere \cite{Tatsuno1982}.  
The net force due to the viscous stress generated by this steady flow on a particle cancels out to zero when the particle has fore-aft symmetry about the $z-$ axis. However, when the particle does not have this symmetry, the steady flow generates a net hydrodynamic force  (torque) that results in a translational (rotational) motion in the levitation plane (see Figure  \ref{fig:Acoustic}a).
The farther the particle density is from the fluid density, the more it will oscillate relative to the fluid, the stronger the steady flow, and thus, the faster it will translate \cite{Wang2012,Nadal2014}.
An interesting aspect of streaming-induced propulsion is that the direction of translation is very sensitive to the rate of asymmetry of the particle, its density distribution, and to the acoustic driving frequency.  Recent experiments \cite{Ahmed2016} and theory \cite{Collis2017} show that a small change in the particle shape, driving frequency or density distribution, can reverse the direction of translation and drastically change the speed  (see \cite{Collis2017} for more detailed explanations on this complex dependence).  

Acoustic levitation can also lead to the spinning of rods and spheres about an axis parallel to the levitation plane   \cite{Wang2012,Zhou2017}. The physical mechanisms behind this behavior are still debated: 
Wang et al.\ \cite{Wang2012} claim that the spinning is related to the helical elastic surface waves (called Rayleigh waves) on the colloids generated by the acoustic field, while, according to Zhou et al.\ \cite{Zhou2017}, the spinning is due to acoustic waves with different phases traveling in the plane that generate a viscous torque on the particles.
Rods have also been observed to orbit in tight circles in the levitation plane \cite{Wang2012,Zhou2017}.
According to \cite{Zhou2017}, these orbiting trajectories come from the torque generated by the steady flow from acoustic streaming on the slightly bent shape of the particles about the rotation axis.

The spinning motion of colloids can be controlled by breaking their rotational symmetry. Sabrina et al.\ engineered flat, twisted-star-shaped particle with fins (see Figure  \ref{fig:Acoustic}e)  \cite{Sabrina2018}. By playing with the parametric configuration of the particle shape, they quantified the relationship between a particle's shape and its rotational motion. In particular, they varied the radius $a$, the rate of chiral asymmetry $c$, and the rotational order $n$ (i.e. the number of fins) of the particles. They showed that the rate of rotation decreases with increasing radius, and changes sign with  chirality, as expected from theoretical works on acoustic streaming. More surprisingly, they observed a non-monotonic dependence of the direction of rotation with $n$ (CW for $n=2,3,6$ and CCW for $n=4,5$). The reversal in the direction of rotation between $n = 3$ and $n = 4$ was predicted by their Boundary Integral numerical simulations. However, the change at $n=6$ is not consistent with their prediction. 
Many effects could explain such discrepancies: neglected high order inertial effects in their simulations and high sensitivity of trajectory with  particle geometry \cite{Collis2017}.

\subsection{Collective behaviour}
\begin{figure}
    \centering
    \includegraphics[width=0.95\columnwidth]{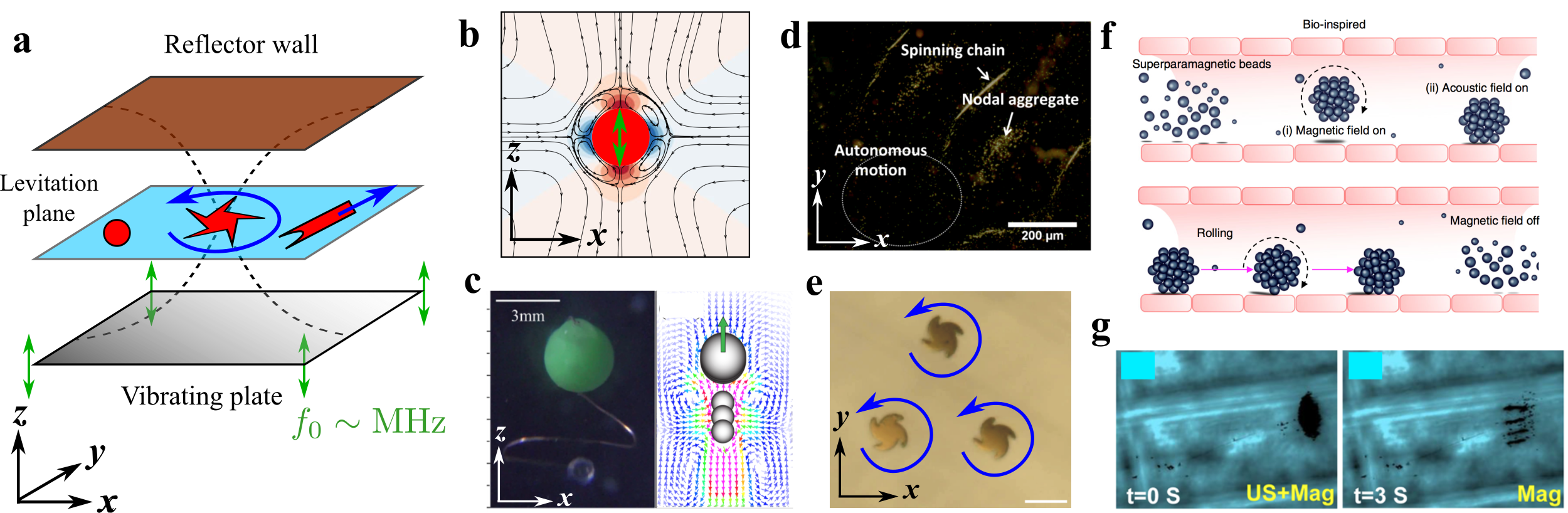}
    \caption{Examples of individual and collective motion in acoustically powered colloidal suspensions. \textbf{a}, Schematic of acoustic levitation and shape-directed motion. 
    \textbf{b}, Streamlines and pressure contours of the acoustic streaming flow induced by an oscillating sphere (simulations), adapted from \cite{Fabre2017}.
    \textbf{c}, Left: Experimental image of a two-sphere swimmer propelled by acoustic streaming \cite{Klotsa2015}. Right: The simulated time-averaged flow field and direction of motion of the swimmer \cite{Jones2018}. \textbf{d}, Experimental image illustrating the formation of spinning chains and aggregation in suspensions rod-shaped acoustically-propelled colloids \cite{Ahmed2016}. \textbf{e}, Experimental image of three co-rotating star-shaped particles, illustrating their equilibrium configuration (scale bar is 15 microns) \cite{Sabrina2018}.  \textbf{f}, The reversible assembly, migration, and rolling  motion of magnetic clusters near boundaries.  These clusters are manipulated with a combination of magnetic and acoustic fields \cite{Ahmed2017}. \textbf{g}, Left: The formation of a static swarm vortex when both magnetic and acoustic fields are on. Right: Directional motion that occurs with the magnetic field only \cite{Li2015}. } 
    \label{fig:Acoustic}
\end{figure}

At the collective level, suspensions of acoustically powered colloids exhibit interesting behaviours due to acoustic radiation, hydrodynamic interactions, or a competition between these two effects. Self assembly of rods and spheres into large chains of particles have been experimentally observed \cite{Wang2012,Zhou2017}. These chains are assumed to form due to secondary acoustic radiation 
forces (i.e.\ Bjerknes forces) that are attractive in the levitation plane \cite{Woodside1997,Groschl1998,Zhou2017}.
When the particles are spinning, the chains form a large vortex that advects particles at large speeds due to hydrodynamic coupling \cite{Wang2012,Zhou2017}.
The acoustic nodal structure in the levitation plane can also generate streak and ring-like structures made of thousands of colloids \cite{Oberti2007,Wang2012}. When the particles are self-propelled, the chains, rings, and streaks translate as well \cite{Wang2012,Zhou2017}.
Zhou et al.\  showed that a suspension of microrods performing orbiting motion could be used to stir  fluid at the microscale\cite{Zhou2017}. Figure  \ref{fig:Acoustic}d shows the formation of spinning chains and nodal aggregates in suspensions of rods \cite{Ahmed2016}.

In addition to secondary acoustic radiative forces, the  hydrodynamic interactions due to acoustic streaming correlate particle motion. \cite{Klotsa2007,Fabre2017} showed that the hydrodynamic forces due to acoustic streaming between two particles leads to an equilibrium configuration that depends on the frequency. \cite{Voth2002} showed that these hydrodynamic effects result in various dynamical states and large scale clustering in suspensions. Recent works have  used the acoustic streaming flow of two unequal spheres connected by a spring to build an acoustically powered swimmer, whose direction of motion depends on the Reynolds number (see Figure  \ref{fig:Acoustic}c) \cite{Klotsa2015,Jones2018}. 

In a standing acoustic wave, co-rotating star-shaped particles repel each other through hydrodynamic repulsion due to acoustic streaming flows, which is in competition with the acoustic confining potentials that push particles into nodal regions. 
As a result, co-rotating spinners form stationary assemblies with a characteristic inter-particle separation distance of approximately 3 particle diameters \cite{Sabrina2018} (see Figure  \ref{fig:Acoustic}e).

Acoustic field have been coupled to other external fields to trigger self-assembly. \cite{Ahmed2014} combined acoustic levitation with magnetic attraction in suspensions of tri-metallic, 2 $\mu$m rods to form motile asters, whose velocity and direction of motion result from asymmetric acoustic streaming. They also showed that these clusters could be guided using an external magnetic field.
Inspired by the migration of leukocytes towards vascular walls and rolling motion before transmigration, \cite{Ahmed2017} combined acoustic and magnetic fields to design bio-inspired magnetic rolling clusters, using suspensions of paramagnetic beads in a channel (see Figure  \ref{fig:Acoustic}f).
 Thanks to an oscillating magnetic field about an axis parallel to the wall, paramagnetic beads  formed rotating compact aggregates made of tens of particles. By locating the pressure nodes of a standing acoustic wave outside the microchannel, the primary radiative force drove these clusters towards the boundaries. Once at the boundaries, the aggregates rolled and translated along the walls, just like leukocytes do. Li et al.\
 developed hybrid magnetic nanoswimmers: the chirality of their helical structure allows them to self-propel in a given direction under a rotating magnetic field, and in the opposite direction due to acoustic streaming \cite{Li2015}. By playing with the relative strength of magnetic and acoustic forcing they controlled the collective motion of these particles:  they observed a static swarm vortex when both the acoustic and the magnetic are on, directional collective motion with just the magnetic field and stable aggregation with the acoustic field only (see Figure  \ref{fig:Acoustic}g).

\subsection{Outlook}
Acoustic waves are bio-compatible and present promising applications for particle sorting, targeted drug delivery and micro-surgery \cite{Rao2015}. The complex interplay between acoustic forces and hydrodynamic interactions leads to emergent behavior in large suspensions. These collective effects could be used for other applications such as pumping and mixing in microfluidic environments.

\section{Modeling the Brownian Dynamics of Colloidal Suspensions}

A major parameter in the dynamics of colloidal suspensions is the Reynolds number, Re,  which compares inertial convective effects to viscous dissipation.  The Reynolds number is defined as follows: Re = $ U a/\nu$, where  $U$ is the typical particle velocity, $a$ is the particle size and $\nu$ is the kinematic viscosity of the fluid.
In most situations, driven colloidal particles live in the realm of low Reynolds number, Re $\sim 10^{-5} - 10^{-2}$, where fluid inertia can be neglected and the fluid reaction to perturbations is instantaneous.

Owing to their typical size and velocities, most colloids are subject to thermal fluctuations and live in a viscous fluid (Re $\approx 0$), where the fluid momentum diffuses much faster than the particles. This means that the fluid velocity relaxes very quickly to a steady solution while the particles hardly move \cite{Balboa2013}.
As a result, hydrodynamic interactions  between particles are long-ranged (typically $~1/r - 1/r^3$), and therefore particle trajectories are highly correlated over long distances compared to the particle size. Thus, the dynamics of driven colloidal suspensions is led by not only by the external drive and thermal fluctuations, but also strongly influenced by hydrodynamic collective effects.

Note that, in some cases, like acoustic streaming, the oscillatory motion of the particles is very fast and the Reynolds number becomes finite, typically Re $\lesssim 1$. In this regime, fluid inertia must be accounted for to describe the physics of the system.
Below we provide a short overview of recently developed numerical methods  that model driven colloidal suspensions at low Reynolds number. An extensive recent review for suspensions at finite Reynolds number is given by Maxey \cite{Maxey2017}. 
Here, we only focus on methods that explicitly solve the \textit{inertia-less} Stokes equations for the \textit{incompressible} fluid phase. We refer the reader to specific reviews on alternative methods, such as Lattice-Boltzmann \cite{Chen1998,Aidun2010}, or particle methods  such as Smooth Particle Hydrodynamics (SPH) \cite{Monaghan2005}, Smooth-Dissipative Particle Dynamics  (S-DPD) \cite{Groot1997,Espanol2003}, Multi-particle collision dynamics (MPCD) \cite{Gompper2009} and Stochastic rotation dynamics (SRD) \cite{Ihle2001}.  

Some of the phenomena presented in the sections above involve tens of thousands, even hundreds of thousands, of  particles that interact through hydrodynamics and other interactions. 
Accounting for all these interactions together with thermal fluctuations in large scale Brownian Dynamics simulations is a challenging task that requires efficient numerical methods.
Below we present the basic equations for the dynamics of colloidal suspensions and briefly enumerate the challenges together with the numerical methods developed to address them. This section does not constitute an exhaustive review of the literature, but rather highlights the major developments that brought significant improvements over the traditional Brownian Dynamics methods.

\subsection{Brownian Dynamics equations and numerical challenges}

Consider a collection of $N$ colloidal particles in a viscous fluid with positions $\bm{x}_i, i=1,..,N$ and orientations $\bm{\theta}_i, i=1,..,N$ given by quaternions. Denote $\bm{q}_i = [\bm{x}_i, \bm{\theta}_i]$ the generalized position of particle $i$, and $\bm{Q} = \lbrace\bm{q}_i, i=1,..,N\rbrace$ the set of position vectors.
 In this regime, the equations of motion for a collection of colloidal particles is given by the overdamped limit of the Langevin equations \cite{Delong2015} 
\begin{equation}
    \frac{d\bm{Q}}{dt} = \bm{\mathcal{M}}\bm{\mathcal{F}} +  \sqrt{2k_BT}\bm{\mathcal{M}}^{1/2}\bm{\mathcal{W}} +
    k_BT\partial_{\bm{Q}}\cdot\bm{\mathcal{M}}
    \label{eq:overdamped}
\end{equation}
where 
$\bm{\mathcal{M}}$ is the $6N\times 6N$ mobility matrix that contains all hydrodynamic interactions between the particles, 
$\bm{\mathcal{F}}$ is the $6N\times 1$ vectors that collects the external generalized forces acting on all the particles,
$\bm{\mathcal{M}}^{1/2}$ is the square root (i.e.\ the Cholesky factor) of the mobility matrix,
and $\bm{\mathcal{W}}$ is a vector of white noise processes.
Note that, due to the absence of inertia in the system, one only needs to solve the temporal evolution of the particle positions rather than the fast degrees of freedoms such as velocities, which allows study of larger time scales.

The first term in the RHS of Eq.\ (\ref{eq:overdamped}) corresponds to the deterministic velocities arising from  external forcing. 
The external forces and torques $\bm{\mathcal{F}}$ can be of various origins, such as gravitational, electro-magnetic (after integrating Maxwell's stress tensor over the particle's surface), and acoustic\footnote{For the sake of conciseness, we only focus on hydrodynamic interactions and refer the reader to reviews and books on the modeling of electro-magnetic interactions \cite{Ghaffari2015,Satoh2017} and acoustic waves \cite{Laurell2014,Ley2017} in fluids and colloidal suspensions.}.
Hydrodynamic interactions between immersed particles (written in the mobility matrix) are obtained by solving Stokes equations with no-slip boundary conditions on particles' surfaces. 
Computing the action of the mobility matrix between immersed colloids can be done either by using analytical results for simple geometries (unbounded  \cite{Rotne1969,Yamakawa1970}, single wall \cite{Swan2007}, spherical container \cite{Aponte2016}), or by using numerical methods, like the Immersed Boundary Method \cite{Peskin2002,Mittal2005} or the Force Coupling Method \cite{Maxey2001,Lomholt2003,Yeo2010b}, that explicitly solve Stokes equations for the fluid phase on a grid and approximate the no-slip boundary conditions on the particles surface via interpolation operators.

The second term in the RHS of Eq.\ (\ref{eq:overdamped}) corresponds to the Brownian velocities. In traditional Brownian Dynamics, $\bm{\mathcal{M}}^{1/2}$  is computed with a Cholesky factorization or using polynomial approximations \cite{Fixman1986, Ando2012}, which yields  a computational cost $O(N^3)$ and $O(N^2)$ respectively, making them poorly scalable to large collections of colloids. Alternatively, Landau and Lifshitz \cite{Landau1959}, proposed a method called Fluctuating Hydrodynamics (FH). FH consists in adding a fluctuating stress tensor into Stokes equations, which, after solving Stokes equations, allows to obtain the particle Brownian velocities without having to compute the square root of the mobility matrix. The  corresponding cost depends on the fluid solver and on the discretization of the particles. FH has been successfully implemented into various numerical methods  for particulate flows with \textit{linear scaling} ($O(N)$) such as the Immersed Boundary Method \cite{Atzberger2007,Atzberger2007b,Atzberger2011,Delong2014,Usabiaga2014}, Distributed Lagrange Multipliers \cite{Sharma2004}, Fluid Particle Dynamics \cite{Tanaka2000}, Finite Element Method \cite{Plunkett2014,DeCorato2016}, Force Coupling Method \cite{Keaveny2014,Delmotte2015}, Boundary Element Method \cite{Bao2017} or Particle Mesh Ewald Method \cite{Fiore2017,Fiore2018}.

The third term in the RHS of Eq.\ (\ref{eq:overdamped}) involves the divergence of the mobility matrix with respect to the particles' positions and orientations. It corresponds to the drift term that appears when taking the overdamped limit of Langevin equations \cite{Fixman1978,Grassia1995,Delong2015}. In traditional Brownian Dynamics this terms is never directly computed but approximated using a midpoint time integration scheme developed by Fixman \cite{Fixman1978}. However this scheme has a high computational cost because it requires solving a resistance problem, i.e.\ inverting the $6N\times6N$ mobility matrix, at each time step. 
To overcome this problem, a new wave of algorithms has emerged in the recent years, among which, the Random Finite Differences (RFD) scheme with various extensions \cite{Delong2014,Sprinkle2017}, and the Drifter-Corrector \cite{Delmotte2015,Fiore2017}. The main advantage of these schemes is that they only incur one or two additional evaluations of the action of the mobility matrix per time step, which makes them competitive compared to more traditional tools.

\subsection{Outlook}

Choosing between the aforementioned methods depends on the shape of the colloids, the geometry of the system considered, and the level of resolution desired. For instance, to simulate the microroller instability, we used the analytic far-field mobility matrix for particles near a no-slip boundary \cite{Swan2007}, together with an iterative method to compute the Brownian velocities \cite{Ando2012} and RFD \cite{Delong2014}, and  with an efficient parallelization on GPU's, allowing us to run twenty thousand time iterations with more than thirty thousands particles in a few hours \cite{Balboa2017,Balboa2017b}. 
The design of efficient numerical methods for Brownian Dynamics is an active area of research. Fluctuating Hydrodynamics is being extended to more and more numerical methods and new time integration schemes are still under development.

\section{Perspectives}

One exciting avenue of recent work is the study of non-spherical/asymmetric colloidal particles.  Significant advances have been made in individual particle fabrication \cite{Keaveny2013,Walker2015, Vach2015,Li2015, Klotsa2015,Gutman2016,Morozov2017,Sabrina2018,Brooks2018,Garcia2018}, and there is still much room to explore the influence of particle shape on  collective dynamics.
Exploring shape asymmetry opens the possibility to engineer particles to achieve a given individual or collective behaviour, such as swarming, mixing, advection, or suspension rheology.  Further possibilities include flow devices which separate/sort particles by shape, shape-influenced flow trajectories, and new kinds of shape-determined dynamic structures.

We also look forward to seeing more work on collective motion in acoustically-powered suspensions; this is a rather young area, and we feel it raises fascinating questions.   For example, an interesting property of the shape-directed microspinners \cite{Sabrina2018} we discussed, is that their direction of rotation depends on their chirality. This should allow for the  study of suspensions of counter-rotating particles in a single standing acoustic wave, something which is not possible with other types of external driving. So far, suspensions of counter-rotating particles at low Reynolds number have only been studied with numerical simulations \cite{Yeo2015,Yeo2016}. These simulations showed that suspensions of counter-rotating particles could phase separate and mix passive particles. An experimental realization of such suspensions would be a major breakthrough.  Another exciting direction are recent studies which have shown that particle density distribution can affect the advection direction and speed of colloidal particles \cite{Ahmed2016,Collis2017}. Collis et al.\  showed that the heaviest parts of a particle have a larger relative motion with respect to the fluid, while the lightest ones move  with the fluid \cite{Collis2017}. Thus, creating colloidal particles with an uneven mass distribution could lead to more efficient self-propulsion and have additional effects on collective hydrodynamic interactions.
Optimizing acoustically-driven suspensions will require the assistance of theory and numerical simulations, as in these systems inertia must taken into account. Numerical tools have been developed at the individual or pair level \cite{Klotsa2007,Muller2012,Hahn2015,Collis2017,Fabre2017,Sabrina2018,Jones2018}, but further development is necessary to simulate suspensions including the effects acoustic streaming, hydrodynamic interactions, and shape asymmetry.

As shown in recent works  \cite{Martinez2015,Martinez2015b,Massana2017a,Driscoll2017,Delmotte2017}, collective effects in rotating suspensions near boundaries can be used for guided transport of passive particles. For instance, carpets of microrollers act as an active boundary layer that generates advective flows near boundaries, while fingers and critters can trap and advect particles at large speeds over long distances \cite{Driscoll2017}. 
All these modes of transport have been studied on flat clean surfaces. However, in many practical situations the surfaces may be dirty, rough, curved \cite{Ahmed2017} or elastic (such as membranes). Additionally, the suspending fluid may contain obstacles or exhibit a non-newtonian rheology. 
Therefore, in order to evaluate the feasibility and applicability of these modes of transport in real-world  biological systems, further studies are necessary. It would be interesting to evaluate the effect of the competition between obstacles and hydrodynamic interactions on the stability of micro-carpets, on the wavelength of the fingering instability and on the robustness of the spontaneous self-assembly into critters \cite{Morin2018,Stoop2017}. It will also be interesting to quantify the effect of fluid visco-elasticity \cite{Patteson2016}, composition (e.g.\ liquid crystals \cite{Lavrentovich2016,Straube2018}), and wall elasticity \cite{Trouilloud2008,Dias2013,Ledesma2013,Boryshpolets2013,Daddi2016,Daddi2016b,Daddi2016c,Daddi2018,Daddi2018b,Rallabandi2018} on the collective motion of driven colloidal suspensions. 

Large scale Brownian Dynamics simulations will be necessary to assist experiments on these questions. 
As shown in the previous section, the last decade has witnessed an  surge in the emergence of new methods to simulate driven colloidal suspensions at large scales, close to the scale of lab experiments. Their computational efficiency surpass the traditional Brownian Dynamics methods that are still currently in use. We hope these promising methods will gain visibility in the near future.

\newpage
\pagenumbering{gobble} 
\begin{enumerate}
    \item outstanding interest \cite{Sabrina2018}: Great example of shape engineering to achieve specific motion using acoustic streaming. We are looking forward to seeing the collective behaviour of rotors in large suspensions.
    
    \item outstanding interest \cite{Driscoll2017}: This work combines experiments, theory and simulations  to study a new fingering hydrodynamic instability in a suspension of microrollers. The fingers pinch off to form  self-sustained stable motile structures called ``critters". They show for the first time that self-assembly of microrollers can be driven by hydrodynamic interactions and offer promising applications for particle transport and mixing in microfluidic systems.
    
    \item outstanding interest \cite{Kokot2018}: While vortex states have been seen in a multitude of active/driven systems, this is one of the only systems in which an emergent vortex state appears without the need for geometric confinement.  This offers new possibilities for studying transport by localized vortices as well as the pinning effects of obstacles.
    
    \item outstanding interest: \cite{Stoop2017, Morin2017} Experimental studies of driven suspensions interacting with disordered obstacles.  Understanding the interaction of particles with complex boundaries is a crucial ingredient for mimicking real-world transport. 
    
     \item outstanding interest: \cite{Brooks2018} Systematic theoretical study of shape effects in ICEP propulsion.  Designing driven suspensions with custom behaviors offers the possibility of new functions for microfludic devices, and the ability to mimic biological systems in new ways.
     
     \item outstanding interest \cite{Fiore2017}: Extension of particle mesh Ewald methods with fluctuations. In this paper, the authors set up a very efficient method, using Fluctutating Hydrodynamics, to simulate colloidal suspensions. The computational efficiency allows them to simulate $O(10^6)$ Brownian particles per second. A high-performance implementation of their code on Graphing Processing Units (GPUs) is available as a plugin for the software package HOOMD-blue (\url{http://glotzerlab.engin.umich.edu/hoomd-blue/index.html}) at the following address: \url{https://github.com/stochasticHydroTools/PSE/}.
    
    \item special interest \cite{Collis2017}: Theory paper that quantifies the effect of particle shape, mass distribution and frequency parameter on the mechanisms of self-propulsion under acoustic streaming with a simple, but relevant, analytical model and numerical simulations.
    
    \item special interest \cite{Ahmed2017}:  This paper nicely combines different mechanisms to mimic the motion of leukocytes in blood channels.  Particle clusters are formed using paramagnetic beads in a rotating magnetic field, wall migration is done using acoustic forces and rolling motion on the walls results from the hydrodynamic coupling between the walls and the rotating clusters. 
    
    \item special interest \cite{Balboa2017}: They use recent numerical schemes for Brownian Dynamics  to simulate the fingering instability in large microroller suspensions. This paper nicely illustrates the capacity offered by recent advances in Brownian Dynamics to simulate and quantitatively reproduce experiments at  the lab  scale. All the codes used in this paper are documented and freely available  at \url{https://github.com/stochasticHydroTools/}.

    \item special interest \cite{Li2015}: Hybrid magneto-acoustic microswimmers whose swimming direction can be reversed with the field used. In addition to the promising applications they offer at the individual level, these swimmers exhibit interesting collective behaviour in the form of localized swarm vortices that can be used for pumping and fluid mixing.

    \item special interest: \cite{Maier2016b}  Self-assembled colloidal wheels represent a unique and high-speed mode of transport.  Additionally, due to their tilt even these simple structures are chiral (have a preferred direction of transport), and thus offer a minimal model system to study non-reciprocal motion.
    
    \item special interest: \cite{Klotsa2015}: This paper introduces a smart utilization of acoustic streaming for self-propulsion. The combination of simulations and experiments gives a clear understanding of the mechanisms at play. It would be interesting to see how these swimmers interact collectively.

    \item special interest: \cite{Morin2018} Experimental and theoretical work  studying the  interaction of colloidal flocks with an external field.  These kinds of studies are crucial to understanding the behavior of these emergent states.  Additionally, this work demonstrates these interactions can be used to design new device functionality such as a spontaneous oscillator.
    
    \item special interest: \cite{Yang2017}  This work demonstrates the fine tunablity of ICEP flows.  By creating colloids with varying surface charge and modulating the conductance of the suspending fluid, a variety of particle clusters, including both chiral and achiral structures, were created.

\end{enumerate}

\newpage

\section{Acknowledgement}
This work was supported primarily by the Materials Research Science and Engineering Center (MRSEC) program of the National Science Foundation under Award Number DMR- 1420073.  Additional support was provided by the Division of Chemical, Bioengineering, Environmental and Transport Systems program of the National Science Foundation under award CBET-1706562.

 \bibliographystyle{elsarticle-num-names}

\bibliography{drivensuspensionsbib}

\end{document}